\documentclass[11pt,a4paper]{article}
\usepackage{amsmath}
\usepackage[dvips]{graphicx}
\newcommand{\bc}{\begin{center}}
\newcommand{\ec}{\end{center}}
\newcommand{\be}{\begin{equation}}
\newcommand{\ee}{\end{equation}} 
\newcommand{\ba}{\begin{eqnarray}}
\newcommand{\ea}{\end{eqnarray}}
 
\begin{document}
\begin{flushright}     
GEF-TH   03/11
\end{flushright}
 \baselineskip 24pt
 \bc {\Large \bf Fourier Optics and Time Evolution\\ of De Broglie Wave Packets}

G. Dillon \\ 
\baselineskip 16pt
{\it Dipartimento di Fisica, Universit\`a di Genova\\
INFN, Sezione di Genova} \ec

 \baselineskip 14pt
 \noindent{\large\bf Abstract:}
 It is shown that, under the conditions of validity of the Fresnel approximation, 
diffraction and interference for a monochromatic wave traveling in the $z$-direction may 
be described in terms  of the spreading in time of the transverse ($x,y$) wave packet. 
The time required for the evolved wave packet to yield identical  patterns as given by 
standard optics corresponds to the time for the quantum to cross the optical apparatus. 
This point of view may provide interesting cues in wave mechanics and quantum physics 
education.

\baselineskip 16pt

\section{Introduction}
In dealing with the time evolution of free De Broglie wave packets one may wonder whether 
optical phenomena as diffraction or interference are related in some way to the spreading 
of the wave function.
Consider, for instance, a plane wave trave\-ling in the $z$-direction and impinging on a 
screen with a thin slit along the $y$-axis. Now think  about the slit as a device 
preparing an ``initial" wave function in the $x$-direction (see lower part of Fig. 1).
This wave function will evolve to yield, at a later time $t$, a probability density 
distribution identical to the diffraction pattern given by classical optics on a screen 
at some distance $z$ from the slit (see upper part of Fig. 1). 

\begin{figure}\bc\includegraphics[width=8.5cm]{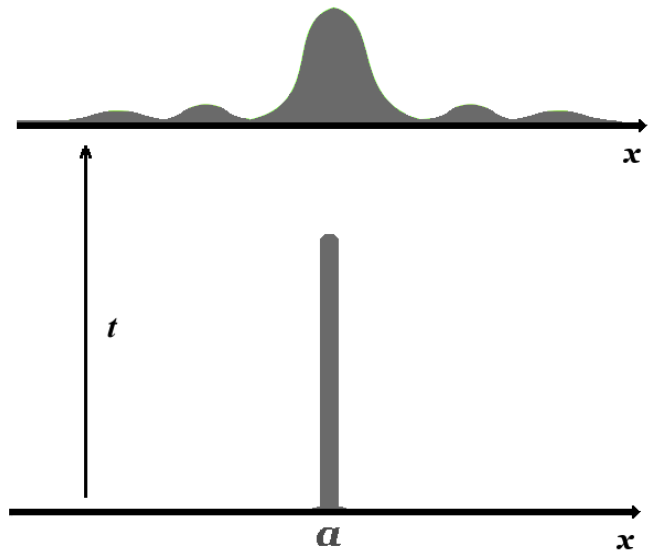}\ec\caption{A rectangular wave 
function, as that created by a slit of width $a$, will evolve to yield, at a later time 
$t$, a probability distribution identical to the diffraction pattern foreseen by 
classical optics on a screen at some distance $z$ from the slit.}
\end{figure}

Analogous considerations are still more instructive in the case of the two-slit 
experiment. 
When the experiment is performed with particles at very low intensities (a particle at a 
time) \cite{Po}, the interference pattern emerges collecting a large number of 
independent events, i.e. performing the same experiment many times under the same 
conditions. The wave mechanical description is simple: Just beyond the screen we get an 
``initial" wave function essentially given by the sum of the two peaks in correspondence 
of the two slits (see lower part of Fig. 2) and properly normalized to 1 (i.e. to one 
particle). At a later time $t$ this wave function will display a probability distribution 
of localization that reproduces the Young interference fringes (see upper part of Fig. 
2).

\begin{figure}\bc\includegraphics[width=8.5cm]{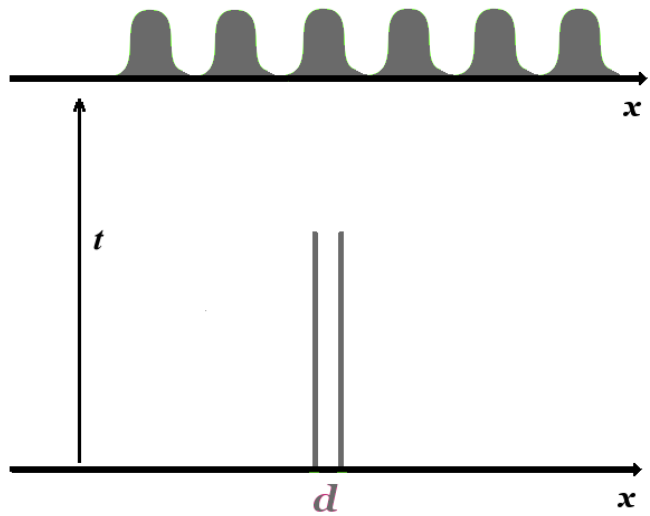}\ec\caption{As in Fig.1 for a wave 
function sum of two real narrow peaks a distant $d$ apart. At a later time $t$ the 
evolved wave function yields a probability distribution that reproduces the Young 
interference fringes.}
\end{figure}

The relation between spreading of one-dimensional wave packets and the diffraction of 
light has  been dealt with in \cite{Van} and recently emphasized \cite{Dho} in connection 
with the observation of nonspreading Airy beams \cite{Siv}. In fact the unusual features 
of these beams were predicted by Berry and Balazs \cite{Be} in the framework of 
one-dimensional wave mechanics. 
Moreover it could provide a helpful tool in teaching quantum physics, especially when 
facing the coherent nature of the wave function,\footnote{In the example of the two-slit 
experiment we have to deal with a coherent superposition of two (nearly) localized 
states; for instance the wave function difference of the same two peaks yields a 
different (shifted) system of fringes and therefore represents a different state.} its 
statistical interpretation, or the subtleties of the two-slit experiment \cite{F}.

Though useful this point of view may be, it does not seem to be exploited for educational 
purposes, as it can be realized by checking existing textbooks on wave mechanics.  The 
reason, I presume, is that the issue still deserves some clarifications. For instance we 
should answer, in a plain way, the following naive questions: We know that the evolution 
of free De Broglie waves for particles of mass $m$ is driven by the {\it dispersion 
relation}: 
\be \omega_k = \frac{\hbar k^2}{2m}
\label{1}\ee
(where $\omega_k$ is the angular frequency for a wave number $k$) which follows 
immediately from the mechanical non-relativistic energy-momentum relation ($E=p^2/2m$) 
and the wave particle relationships: $E=\hbar\omega$ ; $p=\hbar k$. The spreading of De 
Broglie wave packets is known to be due to the dispersive nature of Eq. (\ref{1}). How 
can it be that the final density distributions be identical (apart from the different 
scale of wave lengths) to those of non dispersive ones, like e.m.-waves propagating in 
vacuum? Moreover, does the above evolution time actually correspond to the time elapsed 
for the particle to travel from the slits to the detection screen?

In the present paper the above questions will be addressed.  The time evolved transverse 
wave packet will be compared with the corresponding three-dimensional stationary solution 
derived by the method of the {\it angular spectrum} \cite{G}. As we shall see, the key 
point for matching the two kinds of solution  rests upon the  classical {\it Fresnel 
approximation}. In the framework of this approximation we will find that the time 
required for the wave packet to reproduce the same diffraction or interference pattern as 
that given by the stationary wave, is $t=z/v_g$, where $v_g$ is the group velocity of the 
incident wave and $z$ the distance between the screen with the slits and the detection 
screen. This result is quite general. In the case of experiments performed with 
particles, we could say, using a classical language, it is exactly the time for the 
particle to cover that distance.
\section{Evolution of De Broglie waves}

In this Section we recall briefly the main equations relevant to the time evolution of 
free De Broglie wave packets. Throughout this paper, for simplicity, only one-dimensional 
wave packets will be considered, though the problem at hand applies, in general, to 
two-dimensional wave packets (see next Sect.).

So, let us consider a free particle of mass $m$ on the $x$-axis. If we know the wave 
function $\psi(x,0)$ at the time $t=0$ we get the solution of the (time-dependent) 
Schr\"odinger equation as:
\be
\psi(x,t)=\frac{1}{\sqrt{2\pi}}
	\int_{-\infty}^{+\infty}\hat{\psi}(k)\exp[i(kx-\omega_kt)]dk
	\label{2}
	\ee
where 
\be
\hat{\psi}(k)=\frac{1}{\sqrt{2\pi}}
	\int_{-\infty}^{+\infty}\psi(x,0)\exp(-ikx)dx
	\label{3}
	\ee
	is the Fourier Transform of the initial wave function and $\omega_k$ is given by Eq. 
(\ref{1})

We may write the solution (\ref{2}) in a more compact form by introducing the {\it 
propagator} \cite{S}:

\be
\psi(x,t)=\sqrt{\frac{m}{2\pi i\hbar t}}\int\exp\Big(i\frac{m(x-x')^2}{2\hbar 
t}\Big)\psi(x',0)dx'
\label{prop}
\ee
From Eq. (\ref{prop}) a useful approximation may be drawn when the initial wave packet is 
concentrated in a small region around the origin:
\be
(x-x')^2\approx x^2 - 2xx'
\label{approx}
\ee
that yields:
\be
|\psi(x,t)|^2\approx\frac{m}{2\pi\hbar t}
\Big|\int\exp\big(-i\frac{mx}{\hbar t}x'\big)\psi(x',0)dx'\Big|^2
\label{frau}
\ee
In this approximation the evolved wave function is obtained by simply Fourier 
transforming the initial wave function.

In fact Eq. (\ref{approx}) will be a sensible approximation for the wave packets at hand 
and Eq. (\ref{frau}) will provide a nimble calculation.

 For instance in the case of an initial wave function  different from zero only in a 
small region of width $a$ around the origin and approximately constant therein, we get:
\be
|\psi(x,t)|^2=|N|^2\Big|\int_{-\frac{a}{2}}^{\frac{a}{2}}\exp[-i\frac{mx}
{\hbar t}x']dx'\Big|^2\propto
  \frac{\sin^2(\frac{ma}{2\hbar t}x)}{(\frac{ma}{2\hbar t}x)^2}
  \label{rectangular}
  \ee
  (See Fig. 1).
   
  While in the case of an initial wave function sum of two equal narrow peaks a small 
distance $d$ apart, we can roughly write:
\be
\psi(x,0)=N\Big(\delta(x+d/2)+\delta(x-d/2)\Big)
\label{wf2p}
\ee
and Eq. (\ref{frau}) immediately yields the probability distribution at time $t$:
\be
|\psi(x,t)|^{2}=|N|^{2}\bigg\arrowvert \int_{-\infty}^{\infty}         
\exp[-i\frac{mx}{\hbar t}x']\psi(x',0)dx'\bigg\arrowvert^{2}\propto        
	         \cos^{2}(\frac{md}{2\hbar t}x)
			 \label{2p}
			 \ee
			 (See Fig. 2).
 \section{An outline of the method of the Angular Spectrum}
 In this Section we address the following problem: Consider a monochromatic wave, with 
angular frequency $\omega_0$:
\be  \Phi(x,y,z;t)=\phi(x,y,z)\exp[-i\omega_{0}t] 
\label{Phi}
\ee
incident on the $xy$ plane and traveling in the positive $z$-direction. Suppose we know 
the spatial part of the wave function $\phi(x,y,z)$ at every point of the plane $z=0$; we 
want to calculate the wave function at a subsequent plane $z$.

When (\ref{Phi}) is inserted into the wave equation, one gets the {\em Helmholtz 
equation} for the spatial part :
\be \Delta \phi=-k_0^2\phi
\label{H}\ee
Note that Eq. (\ref{H}) is just the same equation for different kinds of waves: 
Electromagnetic (when described by a scalar function), De Broglie or even Klein-Gordon 
waves; the differences residing in the  relationships between the frequency $\omega_0$ 
and the {\em wave number} $k_0$:

\be
 \omega_{0}= \left\{
\begin{array}{lll}
	  ck_{0} && \textrm{e.m.}
   \\
   \ &\ & \\
   \frac{\hbar k_{0}^{2}}{2m} & &\textrm{De Br.}\\
   \ &\ &\\
   \sqrt{c^2k_0^2+m^2c^4/\hbar^2}& &\textrm{Kl.-G.}
   \end{array}
   \right.
   \label{dr}
   \ee
 
 So, what follows will be true independently of the nature of the waves 
considered.\footnote{Of course for classical waves the wave function is real. So it is 
understood that ultimately the real part of $\Phi$ is to be taken. Defining the classical 
real wave function as: $f=(\Phi+\Phi^*)/\sqrt{2}$ and averaging on the rapidly varying 
factor $\exp[i(k_0z-\omega_0 t)]$ one gets: $\bar{f^2}\equiv|\Phi|^2$.} 
 
 The method proceeds now by Fourier transforming $\phi(x,y,z)$ with respect to the $x,y$ 
coordinates. Since, as anticipated, we assume complete symmetry along the $y$-axis, we 
may get rid of the $y$ coordinate and perform a one-dimensional Fourier transformation:
\be
\phi(x,z)=\frac{1}{\sqrt{2\pi}}\int A(k_x,z)\exp[ik_xx]dk_x
\label{ft}\ee
Then the amplitude $A(k_x,z)$ satisfies:
\be
\frac{d^{2}}{dz^{2}}A(k_x,z)=-(k_{0}^{2}-k_x^{2})A(k_x,z)
\label{eqA}\ee
whose solutions are, in general, combinations of $\exp\big(\pm i\sqrt{k_{0}^{2}-k_x^{2}}\ 
z\big)$.

However, by hypothesis, only progressive waves in the $z$-directions are allowed, so 
that:
\be A(k_x,z)=A(k_x,0)\exp\big(i\sqrt{k_{0}^{2}-k_x^{2}} \ z\big) 
\label{A(k,z)}\ee

The amplitude $A(k_x,0)$ is the Fourier Transform of $\phi(x,0)$:
\be
A(k_x,0)=\frac{1}{\sqrt{2\pi}}\int \phi(x,0)\exp[-ik_xx]dx
 \label{A(k,0)}\ee
 and is known as the {\em angular spectrum} of $\phi(x,0)$ \cite{G}.
 So, from the knowledge of the wave function at the plane $z=0$, we get the wave function 
at the plane $z$:

\be
\phi(x,z)=\frac{1}{\sqrt{2\pi}}\int
	A(k_x,0)\exp[i(k_xx+\sqrt{k_{0}^{2}-k_x^{2}}\, z)]dk_x
	\label{phi(x,z)}
	\ee
 
 Now in the usual optics experiments only small angles intervene, that means the angular 
spectrum $A(k_x,0)$ is different from zero only for $k_x<<k_0$. This is the point where  
the {\em Fresnel approximation} comes through \cite{G}: 
\be\sqrt{k_{0}^{2}-k_x^{2}}\approx k_0-k_x^2/2k_0
\label{FA} \ee
So we get:
\ba \phi(x,z)\approx \frac{\exp(ik_{0}z)}{\sqrt{2\pi}}\int 
	A(k_x,0)\exp[i(k_xx-\frac{k_x^{2}}{2k_{0}}z)] dk_x= \nonumber \\
	 =\exp(ik_{0}z)\sqrt{\frac{k_{0}}{2\pi 
	iz}}\int\exp\Big(i\frac{k_{0}}{2z}(x-x')^{2}\Big)\phi(x',0)dx'
	\label{phiFA}
	\ea
	We can still go further with the {\em Fraunhofer approximation} \cite{G}:
	\be \exp\Big(i\frac{k_{0}}{2z}x'^{2}\Big)\approx 1 
	\label{FHA}\ee
	to obtain
	\be
	\phi(x,z)\propto \int\exp\Big(-i\frac{k_{0}x}{z}x'\Big)\phi(x',0)dx'
	\label{phiFHA}\ee
	Note that Eq. (\ref{FHA}) is  the counterpart of the  approximation (\ref{approx}) in 
wave mechanics.
\vskip 30pt
\section{Matching the two solutions}
As noted, the above equations refer to any kind of waves. For De Broglie waves  we  
compare Eq. (\ref{prop}) with Eq. (\ref{phiFA}). Identifying 
$\psi(x,t=0)\equiv\phi(x,z=0)$ one has:
\be |\psi(x,t)|^2\equiv |\phi(x,z)|^2 
\label{ident}\ee
provided that:
\be t=\frac{mz}{\hbar k_0} \equiv z/v_g
\label{time}
\ee

Eqs. (\ref{ident},\ref{time}) are the main result of the present paper.
Since $k_0$ is the wave number of the incident beam, $v_g=\hbar k_0/m$ is the group 
velocity of the wave traveling in the $z$ direction. According to the small angles 
hypothesis $k_z\approx k_0$,  so  the time (\ref{time}) actually corresponds to the time 
that the particle spends for traveling from the slits to the detection screen. 

As a final point we will check that the same interpretation holds for non-dispersive 
waves as well as for relativistic particles waves.
\begin{itemize}
\item {Electromagnetic Waves}:

The time evolution of a one-dimensional wave packet (of any kind) in a uniform medium is 
formally given by Eq. (\ref{2}), that we rewrite here, for the sake of clarity, with the 
substitutions $k\rightarrow k_x$ and $\omega_k\rightarrow \omega(k_x)$:
\be
\psi(x,t)=\frac{1}{\sqrt{2\pi}}
	\int_{-\infty}^{+\infty}\hat{\psi}(k_x)\exp[i(k_xx-\omega(k_x)t)]dk_x
	\label{2a}
	\ee
	  For De Broglie waves it was of course $\omega(k_x)=\hbar k_x^2/2m$; for  light 
waves, on the other hand, when the Fresnel approximation is valid (i.e. $k_z\approx k_0$) 
one has:
\be
 \omega(k_x)=c\sqrt{k_z^2+k_x^2}\approx ck_0+\frac{ck_x^2}{2k_0}
 \label{drem}
	\ee
So, looking at the dependence of $\omega$ on $k_x$, one sees an {\em effective} 
dispersion relation similar to that of De Broglie waves, with the substitution 
$\hbar/m\rightarrow c/k_0$. This means that these wave packets will spread in time, along 
the $x$ axis, much like De Broglie wave packets. Finally putting 
\be t=z/c\ee in Eq. (\ref{2a}), since $\hat{\psi}(k_x)\equiv A(k_x,0)$, we recover (but a 
phase factor that does not change the diffraction or interference pattern) the stationary 
wave solution $\phi(x,z)$ Eq. (\ref{phiFA})  at the plane $z$.
\item{Klein-Gordon Waves}:

Here (from $E=\sqrt{p^2c^2+m^2c^4}$) we get:
\be
\omega(k_x)=c\sqrt{k_z^2 +k_x^2 +m^2c^2/\hbar^2}\approx c\sqrt{k_0^2 +m^2c^2/\hbar^2}
+\frac{ck_x^2}{2\sqrt{k_0^2 +m^2c^2/\hbar^2}}
\label{drkg}
\ee
to be inserted into Eq. (\ref{2a}).
In this case the matching with the stationary solution Eq. (\ref{phiFA}) is found when:
\be t=z \frac{\sqrt{k_0^2 +m^2c^2/\hbar^2}}{ck_0}=z/v_g
\label{tkg}
\ee
In Eq. (\ref{tkg}) 
\be v_g\equiv\frac{ck_0}{\sqrt{k_0^2 +m^2c^2/\hbar^2}}=pc^2/E
\ee
 is the group velocity of Klein-Gordon waves (at the wave number $k_0$) which coincides 
with the velocity of a relativistic particle with momentum $p=\hbar k_0$.
\end{itemize}
\section{Conclusions}
This paper aimed at clarifying a question that may be exploited in physics education. It 
has been shown that, within the classical Fresnel approximation, the diffraction or 
interference patterns in an optical experiment may be described in terms of time 
evolution of the transverse wave packet. Comparing  the evolved (one- or two-dimensional) 
wave packet at time $t$ with the corresponding ``space evolved" wave function at the 
plane $z$, derived by the methods of Fourier optics, one gets identical patterns when 
$t=z/v_g$ where $v_g$ is the group velocity of the incident wave. Though the main 
interest was focused on De Broglie waves, this result is quite general and holds for any 
waves propagating in a uniform medium. In the case of optical experiments performed with 
particles, we could say that $t$ is the time spent by the particle to cover the distance 
$z$ and this is true for relativistic particles as well.  This point of view may be 
useful, for instance, in  discussing the two-slit experiment performed with single 
electrons.


\begin{thebibliography}{99}
\small
\baselineskip14pt
\bibitem{Po} P. G. Merli, G. F. Missiroli, and G. Pozzi, ``On the optical aspect of 
electons interference phenomena", Am. J. Phys. {\bf 44}, 306-307 (1976); A. Tonomura, J. 
Endo, T. Kawasaki, and H. Ezawa, ``Demonstration of single-electron buildup of an 
interference pattern", Am. J. Phys. {\bf 57}, 117-120 (1989).
\bibitem{Van} G. Vandegrift, ``The diffraction and spreading of a wave packet", Am. J. 
Phys. {\bf 72}, 404-408 (2004).
\bibitem{Dho} K. Dholakia, ``Optics: Against the spread of light", Nature {\bf 451}, 413 
(2008).
\bibitem{Siv} G. A. Siviloglou, J. Broky, A. Dogariu, and D. N.  Christodoulides, 
``Observation of Accelerating  Airy Beams", Phys. Rev. Lett. {\bf 99}, 213901(4) (2007).
\bibitem{Be} M. V. Berry and Balazs, ``Nonspreading wave packets", Am. J. Phys. {\bf 47}, 
264-267 (1979).
\bibitem{F} R. P. Feynman, R. B. Leighton, and M. Sands, {\em The Feynman Lectures on 
Physics} (Addison-Wesley, Menlo Park, CA, 1965), Vol. III.
\bibitem{G} J. W. Goodman, {\em Introduction to Fourier Optics} (Mc Graw Hill Book Co., 
San Francisco, 1968).
\bibitem{S} See for instance: J. J. Sakurai, {\em Modern Quantum Mechanics} (Benjamin 
-Cummings, Menlo Park, CA, 1985).
\end{thebibliography}
\end{document}